\documentclass[pra,openany,oneside,a4paper,12pt,onecolumn,preprint]{revtex4}
\usepackage{amssymb,amsmath,amsthm,bm,dcolumn}
\usepackage[dvips]{graphicx}
\usepackage{color}
\usepackage[english]{babel}
\newcommand{\be}{\begin{equation}}
\newcommand{\ee}{\end{equation}}
\def\tr{\text{tr}}
\def\x{\textbf{x}}
\begin{document}
\title{An initial value representation for the Loschmidt echo}
\author{Eduardo Zambrano}
\author{Alfredo M Ozorio de Almeida}
\affiliation{Centro Brasileiro de Pesquisas F\'\i sicas - CBPF, Rio de Janeiro, RJ, Brazil}

\begin{abstract}
We obtain an initial value representation  for the quantum Loschmidt echo from the semiclassical theory of Wigner function evolution,
together with classical first-order perturbation theory. 
In the limit of small actions, the amplitude of each trajectory reduces to unity, just as in the
\emph{dephasing representation} introduced by Van\'\i\v{c}ek, but these trajectories are here generated by the mean Hamiltonian for both the forward and the backward
motion. This slight change of action may substantially alter the phase. The amplitude correction depends on the second derivative of the action. 
This improved dephasing approximation is verified to work even for quadratic Hamiltonians, for which the semiclassical evolution is exact,
thus extending the range of application beyond its original scope in quantum chaos. 
\end{abstract}

\maketitle
Perturbations are ubiquitous in a wide range of physical systems, so
the study of their consequences on the dynamics is fundamental in both the quantum and the classical realm.
This lies at the heart of the characterization of quantum chaos, since, in contrast to the classical motion, 
the unitary property of quantum evolution 
prevents any progressive separation of quantum states that are initially close. Thus,
an alternative for the characterization of quantum chaos is to study the response to perturbations on the Hamiltonian,
instead of variations on the state itself \cite{Peres}. 
The influence of those perturbations is measured by the celebrated \emph{Loschmidt echo} (LE) or \emph{Fidelity} \cite{Peres,GorinRep}:
  \begin{equation}
  L(t)=\langle\psi|e^{it\hat H_+/\hbar}e^{-it\hat H_-/\hbar}|\psi\rangle,
  \label{LE}
  \end{equation}
  where $\hat H_+=\hat H_-+\delta \hat H$, and $\delta \hat H$ is small compared with $\hat H_-$. 
  This quantity is relevant in quantum information, decoherence \cite{Zurek}, mesoscopic physics \cite{Richter} and in other contexts (for surveys see cf. \cite{GorinRep,Jacquod09,Tomsovic03}),
as well as supporting the notion of `practical irreversibility' in quantum mechanics \cite{Jalabert}.
  Semiclassical theory of the LE has proved a powerful tool to understand its behavior in different regimes,
depending on the duration of the evolution \cite{Jalabert, Bennaker, Tomsovic03, Tomsovic, Prosen,Cucchetti, Wisniacki, Heller, Richter}.
Even so, in practice, this approach usually suffers from the, so called, \emph{root search problem}: The fact that the theory
depends on classical trajectories defined by (hard) boundary conditions, instead of (easy) initial conditions \cite{Tao-Miller}.
This accounts for considerable interest aroused by the surprisingly simple \emph{dephasing representation} (DR) proposed by Van\'\i\v{c}ek \cite{Vanicek04}: 
  \begin{equation}
  \label{DR}
 L_{DR}(t)=\int dx_0 W(x_0)\exp\left(-\frac{i}{\hbar}\int^t_0\delta H(\x(\tau;x_0))d\tau\right).
  \end{equation}
Here $x=(p,q)$ is a point in the $2L$-dimensional phase space, $W(x)$ is the Wigner function of the pure state $\hat\rho=|\psi\rangle\langle\psi|$ and $\x(t;x_0)$ is the classical trajectory of the unperturbed system for the initial condition $x_0$. 
DR has been shown to provide an efficient method to calculate quantum correlation functions, even for higher dimensional systems, 
such as in molecular dynamics \cite{Molecular}; furthermore, DR has also been used to unveil some universal behavior of LE \cite{Wisniacki10}. 
Notwithstanding its practical utility, DR lacks a rigorous deduction. A suggestive argument was proposed in \cite{Vanicek06}, where the \emph{shadowing theorem} of classical mechanics \cite{Shadowing} is invoked. For this reason, the accuracy of DR is commonly associated to the chaotic nature of the system, 
while, so far, its range of applicability is unknown.
  \par
  In this paper we derive an approximation for the LE by evaluating the action in the semiclassical theory for the evolution of the Wigner function within first-order perturbation theory  \cite{OzorioRep,Bohigas}. This approximation reduces to DR for the mean Hamiltonian $\bar H$, if we neglect the semiclassical contribution to the amplitudes. It is important to note that this small change of the Hamiltonian only affects slightly the classical action, but it can lead to significant changes in the semiclassical phase of each trajectory.
Thus, we obtain an estimate of the range of accuracy for the simple original form of DR,
while clarifying which terms are neglected even in the new theory. The focus here is on the case of quadratic Hamiltonians, for which the semiclassical stationary phase method is exact and there is no chaos, though hyperbolicity may be present. In this way, we show that chaotic motion is not necessary for DR to be accurate.
\par
\begin{figure}[htb!]\centering
  \includegraphics[width=9cm]{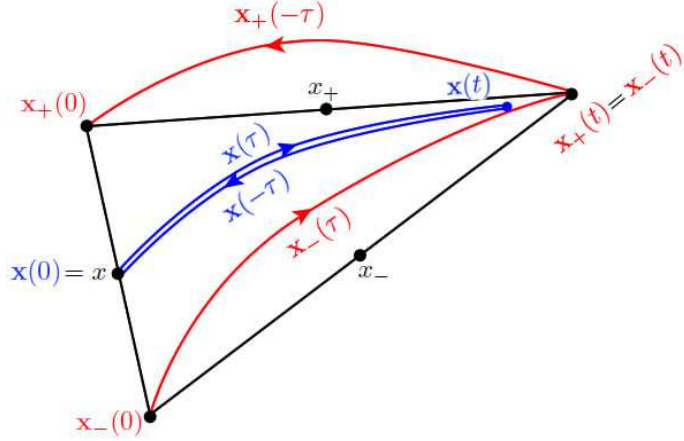}
  \caption{
\label{LE semi fig}Geometrical interpretation of the semiclassical Echo symbol. The blue path is the unperturbed self-retracing trajectory. The red path is the perturbed trajectory from $\x_-(0)$ to $\x_-(0)$, which corresponds to the classical Echo. The black triangle is $\Delta_3(x,x_+,x_-)$.}
  \end{figure}

First, we define the \emph{Echo operator} as  $\hat I_{L}(t)= e^{it\hat H_+/\hbar}\hat I e^{-it\hat H_-/\hbar}$ and $\hat I$ is the identity operator. 
Thus \eqref{LE} can be written as
  \begin{equation}
  \label{LE Weyl}
  L(t)=\tr\,\hat\rho\,\hat I_{L}(t)=\int dx\,W(x)I_{L}(x,t),
  \end{equation}
  where $W(x)$ is the Wigner function of the state $\hat \rho$ and $I_{L}(x,t)$ is the Weyl representation of $\hat I_{L}(t)$. 
  Denoting the Weyl symbol of $e^{-it\hat H_\pm/\hbar} $ by $U^t_\pm(x)$, the Echo symbol is given explicitly by \cite{OzorioRep}:
  \begin{equation}
  \label{echo sym}
  I_{L}(x,t)=\int \frac{dx_+dx_-}{(\pi\hbar)^{2L}}\,U^t_-(x_-)[U^t_+(x_+)]^*e^{\frac{i}{h}\Delta_3(x,x_+,x_-)},
  \end{equation}
where $\Delta_3(x,x_+,x_-)=-2(x_+-x)J(x_--x)$ is the symplectic area of the triangle with middle points $x,$ $x_+$ and $x_-$ (\emph{see} Fig. \ref{LE semi fig}), and $J$ is the $2L\times2L$ standard symplectic matrix. 
The semiclassical expression for the $U_\pm^t(x)$ is then \cite{OzorioRep}
 \begin{equation}
 \label{U semi}
 U_\pm^t(x_\pm)_{SC}=2^L\mathcal{A}_\pm\,\exp\left[\frac{i}\hbar S_\pm^t(x_\pm)\right],
 \end{equation}
where $S^t_\pm$ are the \emph{center} actions, associated to the evolution generated by the classical Hamiltonians $H_\pm(x)$:
 \begin{equation}
  S_\pm^t(x_\pm)=\oint p_\pm\cdot dq_\pm-\int^t_0 H_\pm(\x_\pm(\tau;x_\pm))d\tau.
 \label{center action}
 \end{equation}
Here $\x_\pm(\tau;x_\pm)$ are the classical trajectories for the Hamiltonians $H_\pm$, which are respectively centered on the pair of points $x_\pm$. 
The areas defined by the first integral are closed by the pair of chords that join $\x_\pm(0;x_\pm)$ to $\x_\pm(t;x_\pm)$. 
The amplitudes in \eqref{U semi} are \cite{OzorioRep}
 \begin{equation}
 \mathcal{A}_\pm=\left|\det\left[\mathrm{I}+\mathcal{M}_\pm\right]\right|^{-\frac12}=
  \left|\det\left[\mathrm{I}+\frac12\frac{\partial^2S^t_\pm}{\partial {x_\pm}^2}\right]\right|^{\frac12},
  \label{amplitude}
 \end{equation}
where $\mathrm{I}$ is the $2L\times2L$ identity matrix and $\mathcal{M}_\pm$ are the matrices for the linearized transformations,
$\x_\pm(0;x_\pm)=x_\pm+J\partial S_\pm^t/\partial x_\pm\to \x_\pm(t;x_\pm)=x_\pm-J\partial S_\pm^t/\partial x_\pm$. 
(We have simplified the formulae by omitting the phase space dependence of all amplitudes.)
  Inserting \eqref{U semi} into \eqref{echo sym}, we obtain 
  \begin{equation}
  I_{L}(x,t)_{SC}=
  \left(\frac2{\pi\hbar}\right)^{2L}\int dx_+dx_-\,\mathcal{A}_+\mathcal{A}_-
e^{-\frac{i}{\hbar}\Sigma(x_+,x_-,x)},
  \label{echo sym semi}
  \end{equation}
where $\Sigma=S^t_+(x_+)-S^t_-(x_-)-\Delta_3(x,x_+,x_-)$.
This full semiclassical formula for LE suffers from the need to search for the trajectories centered on each pair of arguments, $x_\pm$.
\par
  The key step is now to reinterpret the above formula for the LE as based on the single classical transformation, which departs from $\x_-(0)$ and arrives at $\x_+(0)$ along the successive trajectories $\x_-(\tau)$, followed by  $\x_+(-t)$ (i.e. the red path in Fig. \ref{LE semi fig}).
Therefore $\hat I_{L}$ is equivalent to an evolution operator associated to this transformation and its Weyl symbol must have the same semiclassical form
as \eqref{U semi}, i.e.
  \begin{equation}
\label{I(x) as propagator}
  I_{L}(x,t)=
  2^L\mathcal{A}_L\exp\left(\frac{i}\hbar S_{L}(x)\right).
  \end{equation}
  Here $S_{L}(x)$ is the center action for the full red trajectory in Fig. \ref{LE semi fig}, evaluated according to \eqref{center action}, whereas $\mathcal{A}_L=\left|\det\left[\mathrm{I}+\mathcal{M}_L\right]\right|^{-\frac12}$, the amplitude for the full transformation $\x_-(0)\to\x_+(0)$, is generated by $S_{L}$ as in \eqref{amplitude}.
The Weyl propagators are unique in that their semiclassical approximation is guaranteed to depend on a single classical trajectory, if it represents an unitary operator in a continuous neighbourhood of the origin \cite{OzorioRep}. This justifies the simple form of
of  \eqref{I(x) as propagator} with no interferences. In contrast, it need not be true for the individual propagators (5), which strictly have superpositions of oscillatory terms \cite{Jalabert}. 
The simplification already attained is then
that a single trajectory centred on $x_-$ must match onto a single trajectory centred on $x_+$ as portrayed in Fig \ref{LE semi fig}. 
\par
Finally, we can consider the red trajectory as resulting from the pair of appropriate perturbations of a single trajectory that is driven by the mean Hamiltonian 
$\bar H= (H_+ + H_-)/2$. That is, $\x(\tau)$ is taken forward during $0<\tau<t$ and backward in $-t< \tau<0$ 
(\emph{see} the double  blue path in Fig. \ref{LE semi fig}),
so that the branches of the full LE trajectory result from $\pm\delta H/2$ perturbations of $\bar H$. 
Then, since the unperturbed trajectory has null action, $\bar{S}_L(x)=0$, the approximation for $S_L(x)$ by first-order perturbation for the round trip  
\cite{Bohigas,OzorioRep} is just
  \begin{equation}
  \label{delta S}
  \delta S_L(x)= -\int_0^t\delta H(\x(\tau;x))d\tau. 
  \end{equation}
  Hence, we obtain the semiclassical LE as
  \begin{equation}
  \label{LE sc}
  L(t)_{SC}=2^L\int dx\,\mathcal{A}_LW(x)\exp\left(\frac{i}{\hbar}\delta S_L(x)\right).
  \end{equation}
  Note that this integral is performed along all the possible initial conditions $x$ for the orbits $\x(\tau)$ of the mean Hamiltonian $\bar H$, thus avoiding the \emph{root search} problem. Since the amplitude $\mathcal{A}_L$ depends on the scalar function $\delta S_L(x)$, it does not require an explicit evaluation of the monodromy matrix, which is computationally expensive for higher dimensional systems \cite{Tao-Miller}. We remark that even though the perturbation is taken along the central orbit for $\bar H$, $\delta S_L(x)$ generates a canonical transformation which is only close to the identity. For small enough perturbations, $\mathcal{M}_L\to \mathrm{I}$, then we recover \eqref{DR}, the DR approximation for the mean Hamiltonian orbit $\x(\tau;x)$. 
  \par
To estimate the error in the action for not using the mean Hamiltonian in the original version of DR, note that the chord,
$\xi(x)\equiv \x_+(0) - \x_-(0)$, is approximated by
  \begin{equation}
  \label{chord} 
 \xi(x) = J\frac{\partial \delta S_L} {\partial x} = -\int_0^t\delta \dot{\x}(\tau;x)d\tau,
  \end{equation}
where $\delta\dot \x$ is the phase space velocity for the Hamiltonian $\delta H(\x)$. The error in the action from the evaluation of \eqref{delta S} along
$\x_-(\tau)$, instead of $\x(\tau)$, is approximately  
 \begin{equation}
  \label{error} 
\delta S_L(x+ \xi/2) - \delta S_L(x) = -\frac{1}{8} \xi\left[ \int_0^t\frac{\partial^2\delta H(\x(\tau;x))}{\partial x^2}d\tau\right]\xi,
  \end{equation}
up to third order terms in the components of $\xi(x)$.
   \par
  Henceforth, we will study in detail the validity of this approximation by comparing it with the usual stationary phase method (SP) applied to \eqref{LE Weyl}. We will restrict to quadratic Hamiltonians, whose semiclassical approximation is exact, because they are the quantum version of (classical) linear canonical transformations \cite{Haake}. A generic quadratic Hamiltonian is
  \begin{equation}
  \label{quadratic H} 
  H(x)=\frac12 x\mathcal{H}x+a\wedge x,
  \end{equation}
  where $\mathcal{H}=\partial^2H/\partial x^2$ is the Hessian matrix and $a\wedge x\equiv aJx$ is the skew product. The associated center generating function for a fixed time $t$
  is given by $S^t(x)=xB^tx+\alpha^t\wedge x,$ where $J B^t=[\mathrm{I}-e^{J\mathcal{H}t}][\mathrm{I}+e^{J\mathcal{H}t}]^{-1}$ and $\alpha^t=2JB^t(J\mathcal{H})^{-1}a$. Note that this relation between $H$ and $S^t$ is not valid when $\mathcal{H}=0$, in which case it reduces to $S^t(x)=-ta\wedge x$. Hence, if we insert the quadratic semiclassical propagators, 
  \begin{equation}
  U_\pm(x)=2^{L}\mathcal{A}_\pm\exp\left(\frac{i}{\hbar}[xB^t_\pm x+\alpha^t_\pm\wedge x]\right),
  \end{equation}
  into the formula \eqref{echo sym}, it remains exact.
\par
\begin{figure}[htb!]
  \centering
  \includegraphics[width=9cm]{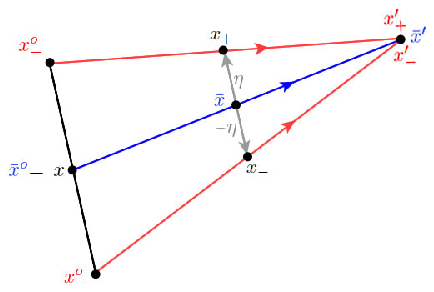}
  \caption{\label{Geom S}
  SP evaluation of LE for the quadratic case. The red arrows correspond to the maps $x_\pm^o\to x'_\pm$, generated by $S_\pm(x_\pm)$, respectively; and the blue chord corresponds to the mean map $\bar x^o=x\to \bar x'$ generated by $\bar S(\bar x)$. These maps have the same final point $\bar x'=x_+'=x_-'$. The middle point between $x_+$ and $x_-$ is $\bar x$, and $2\eta$ is the chord that joins them.}
  \end{figure}

Now, we define the variables $\bar x$ and $\eta$ as $x_\pm=\bar x\mp\eta$ (\emph{see} Fig. \ref{Geom S}), 
  the mean parameters, $\bar B$ and $\bar \alpha$, and their perturbations, $\delta B=B_+-B_-$ and $\delta\alpha=\alpha_+-\alpha_-$, so
  \begin{equation}
  S_\pm(x)=x\left[\Bar B\pm\frac{\delta B}2\right]x+\left(\bar\alpha\pm\frac{\delta\alpha}2\right)\wedge x.
  \end{equation}
  Thus the integral $I_{L}(x,t)$ for quadratic Hamiltonians takes the exact form
  \begin{equation}
  \label{LE bar x eta}
  I_{L}(x,t)=2^L\int d\bar xd\eta\,\mathcal{A}_+\mathcal{A}_-e^{\frac{i}{\hbar}\Phi(\bar x,\eta)}
  \end{equation}
  where $\Phi=\bar x\delta B\bar x+\eta\delta B\eta+4\bar x\Bar B\eta+\delta\alpha\wedge\bar x+2\bar\alpha\wedge\eta+4(\bar x-x)J\eta$.  
  Note that in the general quadratic case the monodromy matrix $\mathcal{M}_L$ is independent of $x$, but $\mathcal{M}_L\neq\mathrm{I}$.
\par The dominant stationary point of the integral with respect to $\bar x$, provides the relation:
  \begin{equation}
  \label{eta}
  \eta=\left[J-\bar B\right]^{-1}\frac{2\delta B\bar x+J\delta\alpha}4\sim\mathcal{O}(\epsilon),
  \end{equation}
  where $\epsilon$ is the size of the perturbation $\delta \alpha$ and/or $\delta B$. Hence, for small $\epsilon$, we may neglect the third-order term $\eta\delta B\eta\sim\mathcal{O}(\epsilon^3)$ in eq. \eqref{LE bar x eta}, 
does not affect condition \eqref{eta}.
  The second stationary phase condition is then
  \begin{equation}
  \label{X no dep eta}
  x\simeq[\mathrm{I}+J\Bar B]\bar x -\frac{\bar\alpha}{2}\text{ or }
  \bar x \simeq [\mathrm{I}+J\Bar B]^{-1}\left(x+\frac{\bar\alpha}2\right).
  \end{equation}
  Thus, in this approximation $x$ does not depend on $\eta$ and it may be interpreted as the initial point of the mean transformation, $\Bar S(\bar x)=\bar x\Bar B \bar x+\bar\alpha\wedge \bar x$, just as in our general derivation of the DR (\emph{see} Fig \ref{Geom S}).
  Combining \eqref{X no dep eta} and \eqref{eta} we have that $4\bar x\Bar B\eta+2\bar\alpha\wedge\eta+4(\bar x-x)J\eta=0$, therefore the SP approximation for the Echo symbol of quadratic Hamiltonians is
  \begin{equation}
  \label{LE sp}
  I^q_L(x,t)=2^L\mathcal{A}_L\exp\left(\frac{i}\hbar\mathcal{S}^q(\bar x,\eta)\right)
  \end{equation}
  where $\mathcal{S}^q(\bar x,\eta)=\bar x\delta B\bar x+\delta\alpha\wedge\bar x$, $\bar x(x)$ is given by \eqref{X no dep eta} and $\mathcal{A}_L$ is the same as in \eqref{LE sc}. 
\par
  This formula has the same form and amplitude as \eqref{LE sc}, but the phase may differ, though we can show that they are equal for the simple case of linear perturbation $\delta H=\delta a\wedge x$: The generic trajectories of a quadratic Hamiltonian \eqref{quadratic H} are 
$\x(\tau)=e^{J\mathcal{H}t}\x(0)+[\mathrm{I}-e^{J\mathcal{H}t}][J\mathcal{H}]^{-1}a,$
  so that
  \begin{equation}
  \label{delta S quadratic}
  \delta S_L(x)=
  \delta a\wedge
[J\mathcal{H}]^{-1}\left([\mathrm{I}-e^{J\mathcal{H}t}](x-[J\mathcal{H}]^{-1}a)-ta\right).
  \end{equation}
  On the other hand, $\bar x=[\mathrm{I}+JB]^{-1}(x+JB[J\mathcal{H}]^{-1}a)$ and so
  \begin{equation}
  \mathcal{S}^q(x)=
  \delta a\wedge
[J\mathcal{H}]^{-1}\left(
[\mathrm{I}-e^{J\mathcal{H}t}](x-[J\mathcal{H}]^{-1}a)+2JB[J\mathcal{H}]^{-1}a\right).
  \end{equation}
  For small times $B\to-t\mathcal{H}/2+\mathcal{O}(t^3)$, thus we recover \eqref{delta S quadratic}. Furthermore, in the simpler case where $\mathcal{H}=0$,
\eqref{delta S quadratic} is always exact; explicitly, 
  $\delta S_L(x)=-t\delta a\wedge\left(x-\frac{ta}2\right)=\mathcal{S}^q(x),$
  because $\bar\alpha=-ta$, $\delta\alpha=-t\delta a$ and $\bar x= x-ta/2$. Finally, we note that here $\delta S_L(x)$ is a linear function, so that there is no correction to the unit amplitude in the original version of DR.
Note that in the foregoing case $\eta\delta B\eta =0$, so that the simplification of \eqref{LE bar x eta} is exact.
  \par
  Another simple example is the Harmonic Oscillator, $\hat H=\omega(\hat p^2+\hat q^2)/2$, so the Hessian is $\mathcal{H}=\omega \mathrm{I}$. We perturb the system by means of two \emph{squeezings}: a contraction in position and an expansion momenta and vice-versa, obtaining the two Hessians
  \begin{equation}
  \mathcal{H_\pm}=
  \omega\mathrm{I}\pm\epsilon\omega\begin{pmatrix}
		    1&0\\0&-1
	  \end{pmatrix}\equiv
  \mathcal{H}\pm\frac{\delta \mathcal{H}}2.
  \end{equation}
  Thus the phase in \eqref{LE sc} is given by
  \begin{equation}
\label{OH dS_L}
  \delta S_L(x)=\frac\epsilon2(q^2 - p^2) \sin(2 \omega t)+
2\epsilon p q \sin^2(\omega t) .
  \end{equation}
  The pair of corresponding symmetric matrices are $\omega B_\pm=-\tan(\frac{\omega t}2)\mathcal{H}_\pm$ \cite{OzorioRep}. Thus $\bar B=-\tan(\frac{\omega t}2) \mathrm{I}$ and
  \begin{equation}
  \delta B=-\tan\left(\frac{\omega t}2\right)\begin{pmatrix}
		    1&0\\0&-1
	  \end{pmatrix}
  \end{equation}
  Then the phase by SP approximation is 
\begin{equation}\mathcal{S}^q(x)=x\left([\mathrm{I}+J\bar B]^{-1}\right)^{T}\delta B[\mathrm{I}+J\bar B]^{-1}x,\end{equation} 
  which 
 is equal to \eqref{OH dS_L}. 
Noted that, in this case the mean Hamiltonian has a discrete spectrum, so that the semiclassical analysis of the different regimes for the decay of the full LE \cite{GorinRep,Tomsovic03,Wisniacki10} can also be carried out.
However, the interval for these must here be much shorter than the period, $T=2\pi/\omega$.
\par
  The last example we consider is the inverted oscillator, which is not chaotic but has hyperbolic dynamics. The Hessian is
  \begin{equation}
  \mathcal{H}=\omega \begin{pmatrix}1&0\\0&-1\end{pmatrix},
  \end{equation}
  and the perturbation $\delta \mathcal{H}=2\epsilon \omega \mathrm{I}$. Thus we have are dealing with the two Hessians $\mathcal{H}_\pm=\mathcal{H}\pm\epsilon\omega \mathrm{I}$. 
  So we obtain that
  \begin{equation}
  \label{delta S inv}
  \delta S_L(x)=
- \frac\epsilon2(q^2 + p^2) \sinh(2 \omega t)
-2 \epsilon p q  \sinh^2(\omega t). 
%
  \end{equation}
  On the other hand $\omega B_\pm=-\tanh\left(\frac{\omega t}2\right)\mathcal{H}_\pm$,
  then $\delta B=-2\epsilon\tanh(\omega t/2)\mathrm{I}$ and $\omega\bar B=-\tanh\left(\frac{\omega t}2\right)\mathcal{H}$
  therefore
  $\mathcal{S}^q(x)$ is equal to \eqref{delta S inv}. 
The determinant associated to the quadratic form \eqref{delta S inv} provides an estimation for dominant contribution to LE, given an initial coherent state. This is found to be independent of the pertubation itself and so LE decays as $\sim e^{-\omega t}$. This `Lyapunov regime' is different from \cite{Jalabert}, where the same factor governs the intensity, not the amplitude in \eqref{LE Weyl}. It should be noted that here we are not performing an average over an incoherent ensemble.
\par
Summarizing, we have obtained a clear derivation of an initial value representation for the LE, which introduces phase and amplitude refinements into the original DR. It was essential to take the evolution for the mean Hamiltonian $\bar H$, in order to obtain the same phase in the above examples. Furthermore, in these cases where the perturbation affects a constant Hessian matrix, the amplitude has a constant second-order correction in $\epsilon$. The object of our theory is the propagation kernel for the full LE, so that appropriate averages over initial states would be required to reconsider the various scenarios of fidelity decay \cite{GorinRep,Tomsovic03,Wisniacki10}. Without such averages, the echo intensity need not decrease monotonically. Indeed, a second parameter in the perturbation allows for isolated zero overlaps, in the simple case of small phase space translations; they can be readily calculated within the present semiclassical framework \cite{ZO-Small}.
 Moreover, the examples above show that the accuracy of the semiclassical approximation for the LE does not require that the motion be chaotic, as long as 
$\eta\delta B\eta$ is small. 
This neglected term can be evaluated within our approximation to provide an estimation of the error in the $L(t)_{SC}$. Evidently, in the general case where
the Hamiltonian is not quadratic, $\delta B$ should be replaced by the Hessian matrix of the center action corresponding to the mean Hamiltonian. 
\par

\par
Financial support from Faperj and CNPq and INCT-IQ is gratefully acknowledged. We thank D. A. Wisniacki, R. Vallejos, F. Toscano, J. Van\'\i\v{c}ek and S. Tomsovic for useful discussions. A.M.O.A.  gratefully acknowledges the kind hospitality of the MPI-PKS, Dresden.

  \end{document}